\documentclass[pre, nofootinbib, floatfix, notitlepage]{revtex4-1}
\usepackage[utf8]{inputenc}

\usepackage{amssymb,amsthm,amsmath, amsfonts}
\usepackage{color, graphicx, enumerate}
\usepackage[font={scriptsize}]{caption}
\usepackage{verbatim}

\begin{document}
\author{Katarzyna Macieszczak
}
\affiliation{University of Nottingham, School of Mathematical Sciences, School of Physics \& Astronomy, University Park, NG7 2RD Nottingham, UK}
\title{The Zeno limit in frequency estimation with non-Markovian environments}
\begin{abstract}
We prove the Zeno limit $N^{3/2}$ of the quantum Fisher information for frequency estimation with a general initial state of $N$ two-level atoms in the presence of local non-Markovian dephasing noise, which was demonstrated for a GHZ state by A. W. Chin, S. F. Huelga, M. B. Plenio (2012). For collective dephasing we prove that non-Markovian noise allows for better scaling for both correlated and uncorelated states. The quantum enhancement in precision is at least as good as in the local case.  \end{abstract}
\maketitle

\textbf{Introduction.} The task of estimating an unknown value of a parameter of quantum system dynamics is cruicial for numerous applications of quantum technologies, e.g. frequency estimation (spectroscopy) in atomic clocks~\cite{AtC}. It is important, however, to be sure that the usually difficult preparation of the system in an entangled state leads to the resolution surpassing the classical limit given by an uncorrelated initial state (the shot-noise limit). 

For parameter estimation such an enhancement was first considered in~\cite{Caves} and for frequency estimation it was first discussed in~\cite{Wineland}. For the system of $N$ two-level atoms with the energy gap $\hbar\omega$, in the case of unitary evolution, the maximally correlated state (the GHZ state) makes it possible to achieve the Heisenberg scaling $\propto N^{-2}$ in resolution, whereas the uncorrelated state gives the shot-noise scaling $N^{-1}$. If, however, there is interaction with an environment, such quantum enhancement in resolution may be signifinatly limited. In~\cite{Huelga} frequency estimation using both the GHZ and the uncorrelated state in the presence of local Markovian dephasing was proved to show the scaling $\propto N^{-1}$. In~\cite{Escher}  this scaling was proved for a general state when $N\rightarrow\infty$ and thus the best possible quantum enhancement was simply a constant. For local non-Markovian dephasing the comparison between GHZ and uncorrelated states was made in~\cite{CHP} for several models of a Gaussian environment. Frequency resolution using the GHZ state was shown to have the new scaling $\propto N^{-\frac{3}{2}}$, whereas the uncorrelated state again gave $\propto N^{-1}$. The new scaling was shown to be due to the quantum Zeno effect~\cite{Zeno}. 

Here we prove that the Zeno scaling is indeed the best possible scaling for a general correlated initial state and is again related to the quantum Zeno effect. We use the bound for phase estimation in the presence of general Gaussian dephasing derived in~\cite{KM}. Furthermore, for collective dephasing we prove that non-Markovian noise allows for better scaling for both correlated and uncorrelated states. Moreover, there is at least the $\propto N^{-1/2}$ quantum enhancement in the resolution scaling for collective non-Markovian dephasing.\\


\textbf{Fisher information.} In frequency estimation with dephasing an unknown value of the frequency $\omega$ is estimated from the result of a POVM measurement $\{\Pi\}_{x\in X}$ performed at time $t$ on an evolved state $\rho_{\omega,t}$.  In one-shot experiment the $\omega$ frequency cannot be estimated perfectly, even when there is no dephasing, due to the projection noise, i.e. observing the experiment result $x\in X$ with the probability distribution $p_\omega(x):=\mathrm{Tr}(\rho_{\omega,t}\Pi_x)$. The best possible resolution is given by the inverse of the Fisher information $F_\omega$: 
\begin{equation}
F_\omega=\int_X\,\mathrm{d}\,p_\omega(x)\left(\frac{\partial}{\partial\omega}\log p_\omega(x)\right)^2,
\end{equation} 
The optimal choice of POVM measurement leads to the quantum Fisher information, which yields the bound for the best resolution in the experiment which uses the initial state $\rho$ and lasts time $t$ ~\cite{QFI}. This can be used to check the performance of a given experiment scheme.

In the presence of dephasing the evolved state $\rho_{\omega,t}$  is determined by the initial state $\rho$, unitary dynamics governed by the Hamiltonian $H=\frac{1}{2}\sum_{j=1}^N\sigma_j^z$, where $\sigma^z$ is the Pauli matrix along the $z$-axis, and dephasing channel $\Lambda_t$: $\rho_{\omega,t}:=e^{-i\omega t H}\Lambda_{t}(\rho)e^{i\omega t H}$, since dephasing commutes with the unitary dynamics. Thus, the quantum Fisher information does not depend on the $\omega$ value and equals:
\begin{equation}
F_{\rho,t}\,=\,t^2\,\mathrm{Tr}\left(\Lambda_t(\rho) L_{\rho,t}^2\right), \qquad\mathrm{where}\qquad \Lambda_t(\rho) L_{\rho,t}+ L_{\rho,t}\Lambda_t(\rho)=-2i\left[H,\Lambda_t(\rho)\right].
\end{equation}

\textbf{Examples of initial states.} For the GHZ state and the uncorrlated CSS state we have (in the eigenbasis of $H$):
\begin{equation}
\rho_{GHZ,t}=\frac{1}{2}\left(\begin{array}{cc}
1& e^{-\gamma(t) N^k}   \\
e^{-\gamma(t) N^k} & 1 \end{array} \right),\qquad\mathrm{and}\qquad \rho_{CSS,t}^{loc}=\frac{1}{2^N}\left(\begin{array}{cc}
1& e^{-\gamma(t)}   \\
e^{-\gamma(t)} & 1 \end{array} \right)^{\otimes N},
\end{equation}
where for the GHZ state $k=1,\,2$ correspond to independent and collective dephasing, respectively. Thus, we obtain $F_{GHZ,t}=N^2 t^2 e^{-2\gamma(t) N^k}$ and $F_{CSS,t}^{loc}=N t^2 e^{-2\gamma(t)}$. $2\gamma(t)$ is the variance of the random phases introduced into the system state due to interaction with a Gaussian environment and depends on the interrogation time $t$. $\gamma(t)$ is determined by the Gaussian environment spectrum, see~\cite{CHP} for examples.

For a general initial state $\rho$ the quantum Fisher information $F_{\rho,t}$ is difficult to calculated, even numerically. Therefore, in order to investigate the asymptotic behaviour of the quantum Fisher infromation we use the upper bound from~\cite{KM} for general dephasing modified to frequency estimation:
\begin{equation}
F_{\rho,t}^{loc} \,\leq\,t^2\, N\,\left(2 \gamma(t)+\frac{N}{I_\rho} \right)^{-1}\qquad\mathrm{and}\qquad F_{\rho,t}^{col}\, \leq\,t^2 \,\left(2 \gamma(t)+\frac{1}{I_\rho}\right )^{-1}, \label{eq:bounds}
\end{equation}
where $I_\rho:=\mathrm{Tr}\left(\rho L_{\rho}^2\right)$ does not depend on $t$ and $\rho L_{\rho}+ L_{\rho}\rho=-2i\left[H,\rho\right]$.  The use of the bound from~\cite{KM} is justified, as when deriving it only the fact that the random phases have Gaussian distribution, and no assumption about the noise Markovianity/non-Markovianity, was used. For the examples given above we have: $I_{CSS}=N$ and $I_{GHZ}=N^2$, which equals $\max_\rho I_\rho$, where $\rho$ is a state of $N$ two-level atoms. \\

\textbf{Frequency estimation.} We consider a total time $T$, which can be divided into an arbitrary number $n$ of individual experiments in which we use the system of $N$ two-level atoms.  If $t_1$, ..., $t_n$ denote the interrogation times of the individual experiments, we have $t_1+...+t_n=T$. Let $\rho_1$,...., $\rho_n$ be the initial states for these experiments. Thus, the best resolution of this series of experiments  is given by the inverse of the sum $F_T:=\sum_{j=1}^n F_{\rho_j,t_j}$. For the fixed total time $T$ we want to find the optimal interrogation times and the initial states to maximise $F_T$.\\

 Let us consider \textbf{Markovian dephasing}. Since the dephasing channel has the semi-group structure: $\Lambda_{t_1+t_2}=\Lambda_{t_1}\Lambda_{t_1}$, we have $\gamma(t)=\gamma t$. Using the bounds in Eq.~(\ref{eq:bounds}) we obtain:
\begin{eqnarray}
F_{T}^{loc} &\leq& N\,\sum_{j=1}^n t_j^2\, \left(2 \gamma t_j+\frac{N}{I_{\rho_j}} \right)^{-1}\leq N\,\sum_{j=1}^n t_j^2\, \left(2 \gamma t_j \right)^{-1} =N\,\frac{T }{2 \gamma } \label{eq:boundMloc}\qquad\mathrm{and}\\
F_{T}^{col} &\leq& \sum_{j=1}^n t_j^2\, \left(2 \gamma t_j+\frac{1}{I_{\rho_j}} \right)^{-1}\leq \sum_{j=1}^n t_j^2\, \left(2 \gamma t_j \right)^{-1} =\frac{T }{2 \gamma }.
\label{eq:boundMcol}
\end{eqnarray}
The bound in Eq.~(\ref{eq:boundMloc}) was proved in~\cite{Escher}. 
We see that for the local noise the best possible scaling of the Fisher information for a fixed total time $T$ is $\propto N$, whereas in the collective case it is constant.

Let us demostrate that, in the case of local Markovian dephasing, for both the GHZ state and the uncorrelated state, the quantum Fisher information scaling  is linear $\propto N$ and thus there is no quantum enhancement is scaling. For the GHZ state we have $F_{GHZ,T}=N^2 T t  e^{-2\gamma t N}$, which attains its maximum $N \,\frac{T}{2\gamma e}$ at $t=(2\gamma N)^{-1} $. For the uncorrelated CSS state we have $F_{CSS,T}=N T t  e^{-\gamma t}$, which has exactly the same maximum value $N \,\frac{T}{2\gamma e}$ at $t=(2\gamma)^{-1}$. These examples were considered in~\cite{Huelga}. \\

Let us discuss \textbf{non-Markovian dephasing}. We do not know the general form of $\gamma(t)$, but for any Gaussian enviroment for small times $t$ we have the quantum Zeno effect leading to $\gamma(t)\approx\frac{\gamma^2}{2} t^2$~\cite{Zeno}. Using the bounds in Eq.~(\ref{eq:bounds}) for a general state we obtain:
\begin{eqnarray}
F_{T}^{loc} &\leq& N\,\sum_{j=1}^n t_j^2\, \left( \gamma^2 t_j^2+\frac{N}{I_{\rho_j}} \right)^{-1}\leq N\,\sum_{j=1}^n t_j^2\, \left( \gamma^2 t_j^2 +\frac{1}{N} \right)^{-1} \quad \mathrm{and}\\
F_{T}^{col}  &\leq& \sum_{j=1}^n t_j^2\, \left( \gamma^2 t_j^2+\frac{1}{I_{\rho_j}} \right)^{-1}\leq \sum_{j=1}^n t_j^2\, \left( \gamma^2 t_j^2 +\frac{1}{N^2} \right)^{-1}.
\end{eqnarray}
Using the Lagrange multipliers, one can show that for every $n$ the uniform division of time is optimal and therefore with $n=T/t$ we arrive at:
\begin{eqnarray}
F_{T}^{loc} &\leq& N\,T t\, \left( \gamma^2 t^2+\frac{1}{N} \right)^{-1}\leq\, N^{\frac{3}{2}}\,\frac{T}{2\gamma}\quad \mathrm{and}\\
F_{T}^{col}  &\leq& \,T t\, \left( \gamma^2 t^2+\frac{1}{N^2} \right)^{-1} \leq\, N\,\frac{T}{2\gamma},
\end{eqnarray}
where the optimal interrogation times $t^{loc}=(\gamma^2 N)^{-\frac{1}{2}}$ and $t^{col}=(\gamma N)^{-1}$, which, as we consider $N\rightarrow\infty$, are indeed within the Zeno effect regime. We see that the best possible scaling of the Fisher information for non-Markovian local dephasing is indeed limited to the Zeno scaling $\propto N^{\frac{3}{2}}$, whereas in the collective case it is limited to the linear scaling $\propto N$.

For the GHZ state in the local case we obtain maximum of the quantum Fisher information $N^{\frac{3}{2}} \frac{T}{ \gamma (2 e)^{\frac{1}{2}}}$ at $t=\gamma^{-1}(2 N)^{-\frac{1}{2}}$. Therefore, we have proved that for frequency estimation in the presence of non-Markovian dephasing the best possible scaling is the Zeno scaling $\propto N^{\frac{3}{2}}$.  The example of the GHZ state for the local dephasing was discussed in~\cite{CHP}. In the collective case we obtain the maximum value $N \frac{T}{ \gamma(2 e)^{\frac{1}{2}}}$ at $t=(\sqrt{2}\gamma N)^{-1}$, which proves the best possible scaling of the quantum Fisher information is linear $\propto N$.\\

In order to determine quantum enhancement for non-Markovian enviroments let us discuss frequency estimation using the uncorrelated state. In the local case, for the uncorrelated state we have  $F_{CSS,T}=N T t  e^{-2\gamma(t)}$, which maximum value $N \,C T$ where C is $\sup_{t>0} t  e^{-2\gamma(t)}$. Therefore, there can be $N^{\frac{1}{2}}$ quantum enhancement in frequency resolution when using an entangled initial state. In the collective case we use the bound in Eq.~(\ref{eq:bounds}).  We have $I_{CSS}=N$. With the assumption $\gamma(t)\approx\frac{\gamma^2}{2} t^2$ we obtain:
\begin{eqnarray}
F_{CSS,T}^{col}  &\leq& T t\, \left( \gamma^2 t^2+\frac{1}{N} \right)^{-1}\leq N^{\frac{1}{2}}\,\frac{T}{2\gamma},
\end{eqnarray}
where the optimal interrogation time $t=\gamma^{-1}(2 N)^{-\frac{1}{2}}$, which for $N\rightarrow\infty$ is within the Zeno effect regime. Therefore, there can be at least $N^{\frac{1}{2}}$ improvement in the Fisher information scaling when using a correlated initial state.\\
 
\begin{figure}[htb!]
\includegraphics[width=0.6\textwidth]{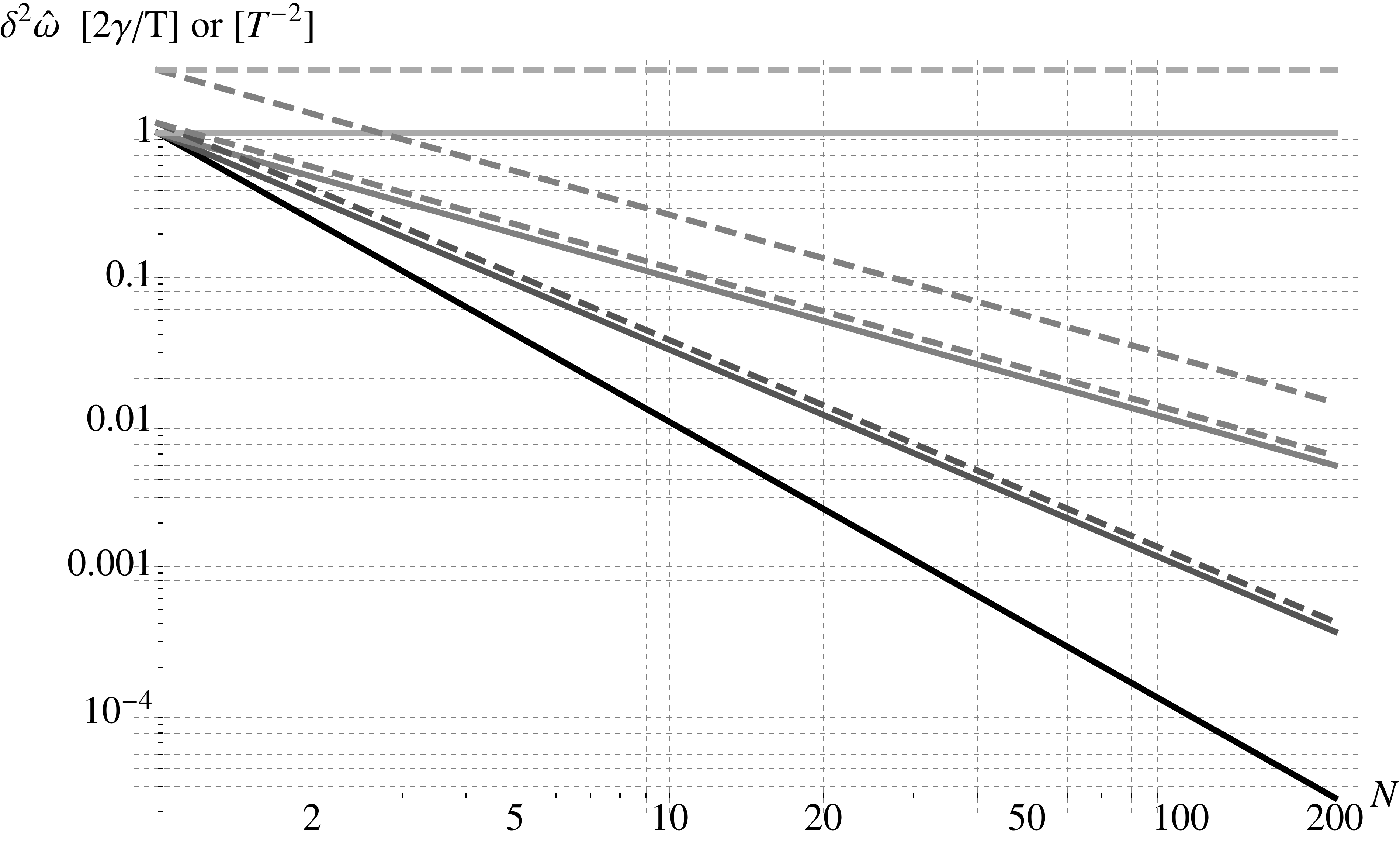}
\caption{Bounds on the frequency estimation resolution for $N$ two-level atoms discussed in the paper. The depicted scalings from the bottom: Heisenberg $(T N)^{-2}$ (unitary, no dephasing), Zeno $\frac{2\gamma}{T} N^{-\frac{3}{2}}$ (local non-Markovian), shot-noise $\frac{2\gamma}{T}  N^{-1}$ (local Markovian \& collective non-Markovian) and constant $\frac{2\gamma}{T} $ (collective Markovian). The dashed lines show the according scaling for the GHZ state. \label{fig:bounds}} 
\end{figure}

\textbf{Summary and comments.} We note that in~\cite{Escher} a tighter bound on the quantum Fisher information in the presence of Markovian independent dephasing was derived:
\begin{equation}
F_{\rho,t}\leq N t^2 \left(e^{2\gamma t} -1 +\frac{1}{N} \right)^{-1} .\label{eq:boundEscher}
\end{equation}
In our view this bound can also be applied to non-Markovian dephasing, replacing $\gamma t\mapsto\gamma(t)$. Since we consider interrogation times $t\rightarrow 0$ when $N\rightarrow\infty$ to derive asymptotic behaviour of the quantum Fisher information for non-Markovian dephasing, the bound in Eq.~(\ref{eq:boundEscher}) is reduced to the bound for $F_{\rho,t}^{loc}$ in Eq.~(\ref{eq:bounds}) used here. Let us note that the bounds in Eq.~(\ref{eq:bounds}) depend on an initial non-dephased state $\rho$, and thus allowing us to consider scaling w.r.t. only the uncorrelated state. 

For a general state of a fixed number of $N$ two-level atoms, one should use the tightest possible bounds when discussing the best experimental startegies. Therefore, it is advisable to use the bound in Eq.~(\ref{eq:boundEscher}) generalised to non-Markovian dephasing. In this paper, however, we focus on the asymptotic behaviour of the quantum Fisher information.

We note that the bound for phase estimation in the presence of Gaussian collective dephasing was first obtained using a variational approach to the Fisher information~\cite{BE}.\\

Here we have proved the Zeno scaling $N^{3/2}$ of the quantum Fisher information for frequency estimation in the presence of local non-Markovian dephasing noise, which was demonstrated for the GHZ state in \cite{CHP}. Furthermore, in the case of collective dephasing, we demonstrate that the constant scaling for Markovian noise ($\propto 1$) is changed to the better scaling  for both correlated ($\propto N$) and uncorrelated states  (at most $\propto N^{\frac{1}{2}}$) when the noise is non-Markovian. In terms of precision this yields at least the quantum enhancement $\propto N^{-\frac{1}{2}}$, which characterises local non-Markovian dephasing. 
\\

\textbf{Acknowledgements.} The author thanks M\u{a}d\u{a}lin Gu\c{t}\u{a}  and Gerardo Adesso for helpful discussions. This research was supported by~the~School of Mathematical Sciences and the School of Physics \& Astronomy at the University of Nottingham.




\begin{thebibliography}{10}
\bibitem{AtC} D. Leibfried \textit{et al.}, Toward Heisenberg-Limited Spectroscopy with Multiparticle Entangled States, \textit{Science} \textbf{304}, 1476 (2004). C.F. Roos \textit{et al.}, 'Designer atoms' for quantum metrology, \textit{Nature} \textbf{443}, 316-319 (2006).
\bibitem{Caves} C. M. Caves, Quantum-mechanical noise in an interferometer, \textit{Phys. Rev. D} \textbf{23}, 1693 (1981).
\bibitem{Wineland} D. J. Wineland \textit{et al.}, Spin squeezing and reduced quantum noise in spectroscopy, \textit{Phys. Rev. A} \textbf{46}, R6797 (1992).
\bibitem{QFI}  S. L. Braunstein and C. M. Caves, Statistical distance and the geometry of quantum states, \textit{Phys. Rev. Lett.}  \textbf{72}, 3439 (1994).
\bibitem{Huelga} S. F. Huelga \textit{et al.}, Improvement of Frequency Standards with Quantum Entanglement, \textit{Phys. Rev. Lett.}  \textbf{79}, 3865 (1997).
\bibitem{Escher} B. M. Escher, R. L. de Matos Filho, L. Davidovich, General framework for estimating the ultimate precision limit in noisy quantum-enhanced metrology, \textit{Nature Physics} \textbf{7}, 406–411 (2011).
\bibitem{CHP} A. W. Chin, S. F. Huelga, M. B. Plenio, Quantum Metrology in Non-Markovian Environments, \textit{Phys. Rev. Lett.} \textbf{109}, 233601 (2012).
\bibitem{KM} K. Macieszczak, Upper bounds on the quantum Fisher Information in the presence of general dephasing, \textit{ArXiv e-Prints} (2014), arXiv:1403.0955 [quant-ph].
\bibitem{Zeno} A. Peres, Zeno paradox in quantum theory, \textit{Am. J. Phys.} \textbf{48}, 931 (1980), O. C. Ghiradi \textit{et al.}, \textit{Il Nuovo Cimento} \textbf{54}, 4 (1979). 
\bibitem{BE} B. M. Escher, L. Davidovich, N. Zagury, and R. L. de Matos Filho, Quantum Metrological Limits via a Variational Approach, \textit{Phys. Rev. Lett.} \textbf{109}, 190404 (2012).
\end{thebibliography}
\end{document}